\documentclass[prb,superscriptaddress,twocolumn]{revtex4-1}
\usepackage[utf8]{inputenc}
\usepackage[T1]{fontenc}
\usepackage{amsmath}
\usepackage{graphicx}
\usepackage{dcolumn}% Align table columns on decimal point
\usepackage{bm}% bold math
\usepackage{epstopdf}
\usepackage{float} 
\usepackage{color}
\usepackage{tikz}
\usepackage{makecell}
\usepackage{comment}
\usepackage{amssymb}
\usepackage[normalem]{ulem}
\usepackage{hyperref}
\usepackage{gensymb}

\def\beq {\begin{equation}}
\def\eeq {\end{equation}}
\def\beqar {\begin{eqnarray}}
\def\eeqar {\end{eqnarray}}
\def\w {\omega}
\def\Im {\textrm{Im}}

\def\bfq {\mathbf{q}}

\def\bfk {\mathbf{k}}

\def\bfe {\mathbf{e}}

\newcommand{\ket}[1]{|#1\rangle}

\newcommand{\hBN}{$h$-BN}

\newcommand{\soleil}{Synchrotron SOLEIL, L'Orme des Merisiers, Saint-Aubin, BP 48, F-91192 Gif-sur-Yvette, France}

\newcommand{\lps}{Universit\'e Paris-Saclay, CNRS, Laboratoire de Physique des Solides, 91405, Orsay, France}
\newcommand{\NewAffWatanabe}{Research Center for Electronic and Optical Materials, National Institute for Materials Science, 1-1 Namiki, Tsukuba 305-0044, Japan}
\newcommand{\NewAffTaniguchi}{Research Center for Materials Nanoarchitectonics, National Institute for Materials Science, 1-1 Namiki, Tsukuba 305-0044, Japan}
\newcommand{\lsi}{LSI, CNRS, CEA/DRF/IRAMIS, \'Ecole Polytechnique, Institut Polytechnique de Paris, F-91120 Palaiseau, France}
\newcommand{\etsf}{European Theoretical Spectroscopy Facility (ETSF)}
\newcommand{\esrf}{European Synchrotron Radiation Facility, 71 Avenue des Martyrs, 38043 Grenoble, France}

\begin{document}

\title{Direct measurement of the longitudinal exciton dispersion in {\hBN} by resonant inelastic x-ray scattering}

\author{Alessandro Nicolaou} 
\email{alessandro.nicolaou@synchrotron-soleil.fr} 
\affiliation{\soleil}%          

\author{Kari Ruotsalainen}
\affiliation{\soleil}%
\affiliation{\esrf}%

\author{Laura Susana}
\affiliation{\lps}%

%\author{Alexander Gloter (?)}
%\affiliation{\lps}%

\author{Victor Por\'{e}e}
\affiliation{\soleil}%

\author{Luiz Galvao Tizei}
\affiliation{\lps}%

\author{Jaakko Koskelo}
\affiliation{\lsi}%
\affiliation{\etsf}%

 \author{Takashi Taniguchi}
\affiliation{\NewAffTaniguchi}%

\author{Kenji Watanabe}
\affiliation{\NewAffWatanabe}%

\author{Alberto Zobelli}
\affiliation{\lps}%
\affiliation{\soleil}%

\author{Matteo Gatti}
\affiliation{\lsi}%
\affiliation{\etsf}%
\affiliation{\soleil}%

\date{\today}

\newcommand{\matteo}[1]{\textcolor{blue}{\textbf{matteo: #1}}}
\newcommand{\alberto}[1]{\textcolor{green}{\textbf{alberto: #1}}}
\newcommand{\kari}[1]{\textcolor{red}{\textbf{kari: #1}}}
\newcommand{\ale}[1]{\textcolor{purple}{\textbf{Ale: #1}}}

\begin{abstract}
We report resonant inelastic X-ray scattering (RIXS) measurements on the prototypical hexagonal boron nitride (\hBN) layered compound.
The RIXS results at the B and N K edges have been  combined with electron energy loss spectroscopy (EELS) experiments and {\it ab initio} calculations within the framework of the  Bethe-Salpeter equation of many-body perturbation theory. 
By means of this tight interplay of different spectroscopies, the lowest  longitudinal exciton of {\hBN} has been identified. Moreover, its qualitatively different dispersions 
along the $\Gamma$K and the $\Gamma$M directions of the Brillouin zone have been determined.
Our study advocates soft x-ray RIXS and EELS to be a promising combination to investigate electronic excitations in materials.

\end{abstract}

%\pacs{}
\maketitle

\section{Introduction}

Resonant inelastic X-ray scattering (RIXS) is a powerful spectroscopy to characterize neutral low-energy excitations in materials \cite{Schulke2007,Kotani2001,Rueff2010,Ament2011}. 
The process of X-ray photon scattering can be understood as the coherent combination of X-ray absorption and X-ray emission. In the simplest situation, a core electron is promoted into an empty conduction state by absorbing a photon, while the core hole is filled by the decay of a valence electron that emits a photon of lower energy. 
The energy lost by the photon matches the energy difference between the ground state and an excited state with an electron in a conduction state and a hole in a valence state.
The resonant character is due to the fact that the incoming photon energy is close to the energy of an X-ray absorption edge.
Being a photon-in photon-out technique, RIXS can be considered as a bulk probe, while its resonant nature makes it element and orbital selective. 
These favorable features, together with the impressive improvements in energy resolution obtained in the recent years\cite{Rueff2010,Ament2011}, have made RIXS a very appealing technique.
Indeed, RIXS has been applied  to succesfully investigate
the elementary excitations of the various degrees of freedom (charge, orbital, spin and lattice) in a wide  range of materials, notably transition metal compounds 
\cite{Hepting2018,Gretarsson2019,Braicovich2009,LeTacon2011,Schlappa2012,Hill2008,Braicovich2010,Lee2013,Lee2014,Moser2015,Rossi2019,PhysRevB.103.235158_Kari_RIXS}. 

However, the richness of RIXS spectra also comes at the price of the difficulty of its theoretical interpretation, which is largely due to the inherently complicated nature of the scattering process, of second order in the photon-matter interaction. An accurate description of the scattering
requires the Kramers-Heisenberg formula
 \cite{Schulke2007,Kotani2001,Rueff2010,Ament2011} for the double differential cross section that describes RIXS spectra $I(\w_1\bfk_1\hat{\bfe}_1;\w_2\bfk_2\hat{\bfe}_2)$ as a function of the energies, wavevectors, and polarizations of the 
incoming and outgoing photons\footnote{Atomic units are used throughout the paper if not otherwise stated.}, respectively.
The {\it ab initio} calculation of RIXS spectra using Green's function theory, within the framework of the Bethe-Salpeter equation\cite{Onida2002} (BSE) in the GW approximation\cite{Hedin1965} (GWA),
has been pioneered by Shirley and coworkers \cite{Shirley1998,Carlisle1999,Carlisle1999b,Shirley2000,Shirley2000b}  in the late nineties. Despite recent progress\cite{Gilmore2015,Vinson2017,Gilmore2021,Dashwood2021,Vorwerk2020,deGroot2021}, its use remains limited.
The main reason is the inadequacy of the first-order perturbative GWA and static BSE standard implementations for transition metal compounds \cite{Faleev2004,Bruneval2014,Cudazzo2020}.
Instead, the state-of-the-art theoretical simulation of RIXS is nowadays mostly grounded on finite cluster models within multiplet ligand-field theory \cite{DeGroot2008,Haverkort2012,Josefsson2012,Zimmermann2018,deGroot2021}. These theoretical methods have repeatedly demonstrated to be very efficient in the description of measured spectra, and have also fixed the current terminology for transition metal compounds \cite{Zaanen1985}.
However, these model approaches are not parameter free,
treat solid state effects with drastic approximations
 
and are limited by their high computational cost to deal explicitly  only with few degrees of freedom.
Therefore, progress in the interpretation of RIXS spectra  still remains a pressing need in order to realise entirely the potential of such a powerful technique. 

A promising strategy to improve our understanding is to establish a careful connection of RIXS with alternative spectroscopy techniques that also measure neutral excitations in materials\cite{Huotari2012}, such as optical absorption, electron energy loss spectroscopy \cite{Fink2001,GarciaAbajo2010,Egerton2011} (EELS), and non-resonant inelastic X-ray scattering  \cite{Schulke2007} (NRIXS).  
The electronic excitation spectra measured by these  complementary methods can be all directly expressed in terms of the macroscoscopic dielectric function\cite{Onida2002} $\epsilon(\bfq,\w)$.
While absorption spectra are described by $\Im \epsilon(\bfq\rightarrow0,\w)$,
both EELS and NRIXS give access to the loss function $-\Im \epsilon^{-1}(\bfq,\w)$.
Optical spectroscopies often have an excellent energy resolution, 
but probe the long-wavelength limit $\bfq\rightarrow0$ only. Conversely, EELS and NRIXS can  measure the momentum dependence of different excitations in solids\cite{Schulke2007,Waidmann2000,Galambosi2005,Cai2006,Larson2007,Abbamonte2008,Huotari2009,Huotari2010,Huotari2011}, typically in the low ($q < 1$ \AA$^{-1}$) and high ($q > 1$ \AA$^{-1}$) momentum ranges, respectively\cite{Huotari2011}. 

When performed within a transmission electron microscope, EELS provides an excellent spatial resolution, down to the atomic scale when looking at core excitations,  allowing one  to correlate structure and spectroscopic response over a wide energy domain.
On the other hand, since it makes use of hard X-rays, NRIXS is compatible with complex sample environments such as high-pressure cells, enabling  pressure-dependent studies of electronic excitations.\cite{Rueff2010,Mao2011} 
Being first-order in the probe-matter interaction, the theoretical interpretation of these spectroscopies is simpler than for RIXS. 
The {\it ab initio} GW-BSE  represents the state-of-the-art approach to characterise excitons\cite{Knox1963} (i.e., bound electron-hole pairs) in the absorption spectra of solids\cite{Onida2002,Bechstedt2014,Martin2016}. Similarly, the GW-BSE and the computationally more efficient approximations of time-dependent density functional theory\cite{Onida2002,Runge1984,Ullrich2012} have been successfully employed to describe electronic excitations  such as plasmons\cite{Pines1963} (i.e., collective excitations of the electronic charge) in the loss functions of a large variety of materials\cite{Quong1993,Aryasetiawan1994,Larson1996,Marinopoulos2002,Gurtubay2005,Weissker2006,Hambach2008,Cazzaniga2011,Echeverry2012,Zubizarreta2014,Seidu2018,Iori2012,Roedl2017,Ruotsalainen2021,Caliebe2000,Soininen2000,Olevano2001,Gatti2013}. 

Here we show how RIXS is a powerful technique for probing the %interesting 
intriguing exciton physics of a prototypical wide-gap $sp$ layered compound, hexagonal boron nitride ({\hBN}). 
{\hBN} is the ionic analogue of graphite\cite{Bassani1975}.
It is formed by weakly interacting two-dimensional hexagonal BN sheets that are stacked along the crystallographic $c$ axis.
Being a material presenting a very strong luminescence in the deep ultraviolet\cite{Watanabe2004,Watanabe2009,Gil2020,Moon2022}, the electronic and optical properties of {\hBN} have been extensively studied  with different experimental techniques, including RIXS\cite{OBrien1993,Jia1996,Carlisle1999,Yanagihara1997,Miyata2002,MacNaughton2005,Vinson2017,Pelliciari2024},
optical spectroscopies (ellipsometry, reflectivity,  photoluminescence or cathodoluminescence) \cite{Mamy1981,Artus2021,Watanabe2004,Watanabe2009,Museur2011,Bourrellier2014,Cassabois2016,Schue2016,Elias2021,Schue2019}, 
EELS\cite{Buechner1977,Tarrio1989,Fossard2017,Schuster2018} and NRIXS\cite{Galambosi2011,Fugallo2015}.
Due to the important role  played by defects within the band gap\cite{Tran2016,Caldwell2019}, the availability of high quality {\hBN} samples\cite{Taniguchi2007} has been of utmost importance in order to obtain reliable results in those experiments.
On the other side, {\it ab initio} calculations have demonstrated that {\hBN} has an indirect band gap and a large exciton binding energy \cite{Catellani1987,Blase1995,Cappellini2001,Arnaud2006,Wirtz2005,Wirtz2006}.
Its simple band structure can be also accurately parametrised by tight-binding approaches\cite{Doni1969,Bassani1975,Galvani2016,Paleari2018}.
Because of the large electronegativity difference  between boron and nitrogen, the BN covalent $sp^2$ bond is polar.
The electronic states can be classified according to the even ($\sigma$) and odd ($\pi$) parity with respect to the BN layer. 
The occupied $\pi$ states are mainly localised on N atoms, while the unoccupied $\pi^*$ mostly on the B ones. The lowest energy electronic excitations are determined by $\pi$-$\pi^*$ transitions that in optical spectra are visible mainly for in-plane polarization.

In the present work, we have measured by RIXS the neutral electronic excitations in a {\hBN} single crystal in the small $q$ range $<$ 0.1 \AA$^{-1}$.
With respect to previous RIXS experiments\cite{Jia1996,Carlisle1999} that focused mostly on band properties scanning the excitation energy in a wide range across $\pi^*$ and $\sigma^*$ resonances, here, by performing RIXS at the B K edge, we have been able to identify the lowest energy exciton of {\hBN} and to %accurately 
track its dispersion.
Moreover, we have combined RIXS with EELS and {\it ab initio} BSE calculations, together with recent ellipsometry data from Ref. \onlinecite{Artus2021}, to establish a firm connection between the different spectroscopies, which has a general validity well beyond the specific case of {\hBN}.

The paper is organised as follows.
Sec. \ref{sec:exp_comp_details} summarizes the experimental and computational details, along with the key definitions of the spectral quantities that are measured and analysed in the rest of the paper.
Secs. \ref{sec:XAS} and \ref{sec:RIXS} give an overview of the XAS and RIXS spectra measured at the B and N K edges.
Sec. \ref{sec:rel_EELS_abs} compares RIXS spectra with EELS and absorption,  also supported by {\it ab initio} calculations, defining the relation between the different techniques.
Sec. \ref{sec:exc_disp} focuses on the exciton dispersion as a function of the wavevector $\bfq$, while Sec. \ref{sec:analogy} suggests an analogy with orbital excitations in transition metal oxides. 
Finally, conclusions are drawn in Sec. \ref{sec:conclusion}.

\section{Experimental and computational details}
\label{sec:exp_comp_details}

\subsection{RIXS at the B and N K edges}
\label{sec:exp_comp_details_RIXS}

RIXS experiments were performed at the SEXTANTS beamline of the synchrotron SOLEIL \cite{SACCHI2013} employing the AERHA spectrometer \cite{ CHIUZBUAIAN2014} at a scattering angle of 85$^\circ$.
High-quality colorless and transparent single crystals of  {\hBN}  were produced by a high-pressure and high-temperature method using a barium-related solvent system as reported in \cite{Taniguchi2007}. Samples were cleaved in air by means of scotch tape under an optical microscope.
They were introduced in the load lock a few minutes after cleaving. 
Two different single crystals were mounted with the $\Gamma$K and $\Gamma$K directions contained in the scattering plane for the measure of the exciton dispersion.  The  \textit{q} resolution of RIXS measurements was better than 0.002 {\AA}$^{-1}$, while the overall energy resolution of RIXS spectra at B and N K edges was 80 and 110 meV full-width-at-half-maximum (FWHM) respectively. 
XAS spectra and RIXS maps were done at room temperature, while the exciton dispersion was studied at 23 K. The experimental geometry is presented in figure \ref{XAS}(a).

\subsection{EELS measurements}
\label{sec:exp_comp_details_EELS}

The same type of high-quality {\hBN} single crystals used for RIXS  were employed for EELS experiments. Microscopy samples were prepared by chemical exfoliation in ethanol of macroscopic {\hBN} crystals, later deposited onto a lacey carbon copper TEM grid to be analyzed.
Monochromated low-loss EELS was performed on a modified Nion HERMES-S200 STEM microscope operated at 60 keV with the electron beam monochromated to an FWHM of 30 meV. Both a convergence and collecting semiangle of 5 mrad were used, resulting in a probe size of approximately 1 nm.
This corresponds to an integrated momentum transfer of about 0.6 {\AA}$^{-1}$ which is about the half of the {\hBN} first Brillouin zone. The sample  was cooled at $\approx$ 100 K and carefully aligned with the [001] crystallographic axis oriented along the electron beam propagation direction.
ELNES measurements were performed with a non-monochromated electron beam.

\subsection{Theory}
\label{theory}

In the GWA with a statically screened Coulomb interaction $W$, the BSE can be cast into an excitonic hamiltonian\cite{Onida2002,Bechstedt2014,Martin2016}: 
\begin{equation}
    \hat H_{\rm exc} = \hat H_{\rm h} + \hat H_{\rm e} + \hat V - \hat W, 
    \label{eq:BSE}
    \end{equation}
    where $\hat H_{\rm h}$ and $\hat H_{\rm e}$  describe the  hole and electron independent propagation, respectively, whereas $\hat V$ is the repulsive electron-hole exchange interaction due to the Coulomb interaction and $-\hat W$ is the attractive electron-hole direct interaction due to the screened Coulomb interaction.
While the latter is responsible for the formation of bound excitons,  $\hat V$ takes into account local-field effects due to electronic inhomogeneities\cite{Adler1962,Wiser1963}. Moreover, its long-range component is at the origin of plasmons\cite{Pines1963}.

RIXS spectra are described by the Kramers Heisenberg equation\cite{Schulke2007,Kotani2001,Rueff2010,Ament2011}, which expresses  the second-order perturbative contribution of the electron-photon interaction $\mathbf{A} \cdot \mathbf{p}$, i.e., the product between the
vector potential $\mathbf{A}$ of the electromagnetic field and the electron momentum $\mathbf{p}$.
When the incident photon energy $\w_1$ is tuned to the vicinity of an absorption edge, it reads:
\begin{widetext}
\begin{equation}
\label{eq:RIXS}
I(\w_1\bfk_1\hat{\bfe}_1;\w_2\bfk_2\hat{\bfe}_2) \propto \sum_f
\left|
\sum_n
\sum_{jj'}^N
\frac{
\left\langle
f \left|
(\boldsymbol\epsilon^*_2\cdot\mathbf{p}_j)e^{-i\mathbf{k}_2\cdot\mathbf{r}_j})
\right|n
\right\rangle
\left\langle 
n\left|
(\boldsymbol\epsilon_1\cdot\mathbf{p}_{j'})e^{i\mathbf{k}_1\cdot\mathbf{r}_{j'}})
\right|g\right\rangle
}
{E_g-E_n+\omega_1-i\Gamma_n/2}\right|^2
\times
\delta(E_g-E_f+\omega),
\end{equation}
\end{widetext}
where $\ket{g}$,  $\ket{f}$ and $\ket{n}$  of energies $E_g$,  $E_f$, and $E_n$ (and lifetime $1/\Gamma_n$), denote, respectively, the ground state, the final state, and an intermediate state of the system of N interacting electrons (located at positions $\mathbf{r}_j$).
Being $\w_2$ the emitted photon energy that is measured in the experiment, the energy conservation implies that the energy loss $\w=\w_1-\w_2$ is equal to the excitation energy $E_f-E_g$ of the electronic system.
%\matteo{explain what is measured and redefine $\w_2$}

Instead, NRIXS spectra, arising from the $\mathbf{A}^2$ electron-photon perturbation at the first order,
measure the dynamical structure factor, defined as\cite{Schulke2007,Rueff2010}:
\begin{equation}
\label{eq:NRIXS}
S(\bfq,\w) = \sum_f\left|\langle f|\sum_j^N e^{i\mathbf{q}\cdot\mathbf{r_j}}|g\rangle\right|^2 \times \delta(E_g-E_f+\omega),
\end{equation}
where the momentum transfer  $\bfq=\mathbf{k}_1-\mathbf{k}_2$.
The dynamical structure factor is proportional to the loss function\cite{Schulke2007}: $S(\bfq,\w) \propto   -q^2\text{Im} \epsilon^{-1}(\bfq,\w)$, where
\begin{equation}
%S(\bfq,\w) \propto  
-\text{Im} \epsilon^{-1}(\bfq,\w) 
=  \frac{\text{Im} \epsilon(\bfq,\w)}{\left[\text{Re} \epsilon(\bfq,\w)\right]^2+\left[\text{Im} \epsilon(\bfq,\w)\right]^2}.
\label{eq:sqw}
\end{equation}
The loss function is also directly measured in EELS \cite{Fink2001,Egerton2011}:
\begin{equation}
\label{eq:EELS}
\textrm{EELS}(\bfq,\w) \propto  -\frac{1}{q^2} \text{Im} \epsilon^{-1}(\bfq,\w).
\end{equation}
Peaks in the loss function $-\text{Im} \epsilon^{-1}$ are at higher energies than peaks in the corresponding absorption spectrum $\text{Im} \epsilon$, since the difference between the two quantities is due to the repulsive long-range component of the Coulomb interaction\cite{Onida2002,Sottile2005}.
In particular, in the loss function collective excitations such as plasmons correspond to zeroes of $\text{Re} \epsilon$ where $\text{Im} \epsilon$ is not too large.

For core levels, absorption, EELS and NRIXS take different names\cite{Huotari2012}, being called: X-ray absorption spectroscopy (XAS), energy loss near edge structure (ELNES), and X-ray Raman scattering (XRS), respectively. 
When the photon energy is near an absorption threshold, XAS is also known as X-ray absorption near edge structure (XANES) or near edge X-ray absorption fine structure (NEXAFS).
In this regime, where the electronic excitations are localised around the atomic sites and the long-range component of $\hat V$ becomes negligible, $\text{Re} \epsilon \sim 1$: the weaker $\text{Im} \epsilon$ spectrum measured by XAS becomes equal to the loss function $-\text{Im} \epsilon^{-1}$ measured by ELNES and XRS in the $\bfq\rightarrow0$ limit.

\subsection{Computational details}
\label{compdetails}

In the present work, BSE calculations followed the computational scheme already successfully employed in  previous works\cite{Fugallo2015,Koskelo2017} on {\hBN}.
We have used Troullier-Martins norm-conserving pseudopotentials\cite{Troullier1991}.
Ground-state calculations in the local density approximation, adopting the experimental crystal structure\cite{Solozhenko1995},  converged with 30 Hartree cutoff energy.
The calculation of the screened Coulomb interaction included 300 bands.
We have used a scissor correction of  2.35 eV and a valence band stretching of 5\% to mimic the difference between LDA and GW band structures\cite{Artus2021}. 
BSE calculations included 8 occupied and 12 empty bands in a 24$\times$24$\times$6 grid of $\bfk$ points (amounting to 3456 inequivalent $\bfk$ points in the Brillouin zone) and with a cutoff of 8.3 Hartree for the e-h interactions.
Confirming the conclusions of Ref. \onlinecite{Olevano2001}, we have found that the 
inclusion of the contribution  from antiresonant transitions is necessary to obtain an accurate loss function, while the coupling between resonant and antiresonant transition has negligible effects.
We have used \texttt{Abinit} \cite{Gonze2016} for the ground-state calculations and  \texttt{Exc} \cite{EXCcode} for the BSE simulations of loss spectra.

\section{Results and analysis}

\subsection{XAS and ELNES spectra}
\label{sec:XAS}
	
\begin{figure}[t]
    \includegraphics[width=0.8\columnwidth]{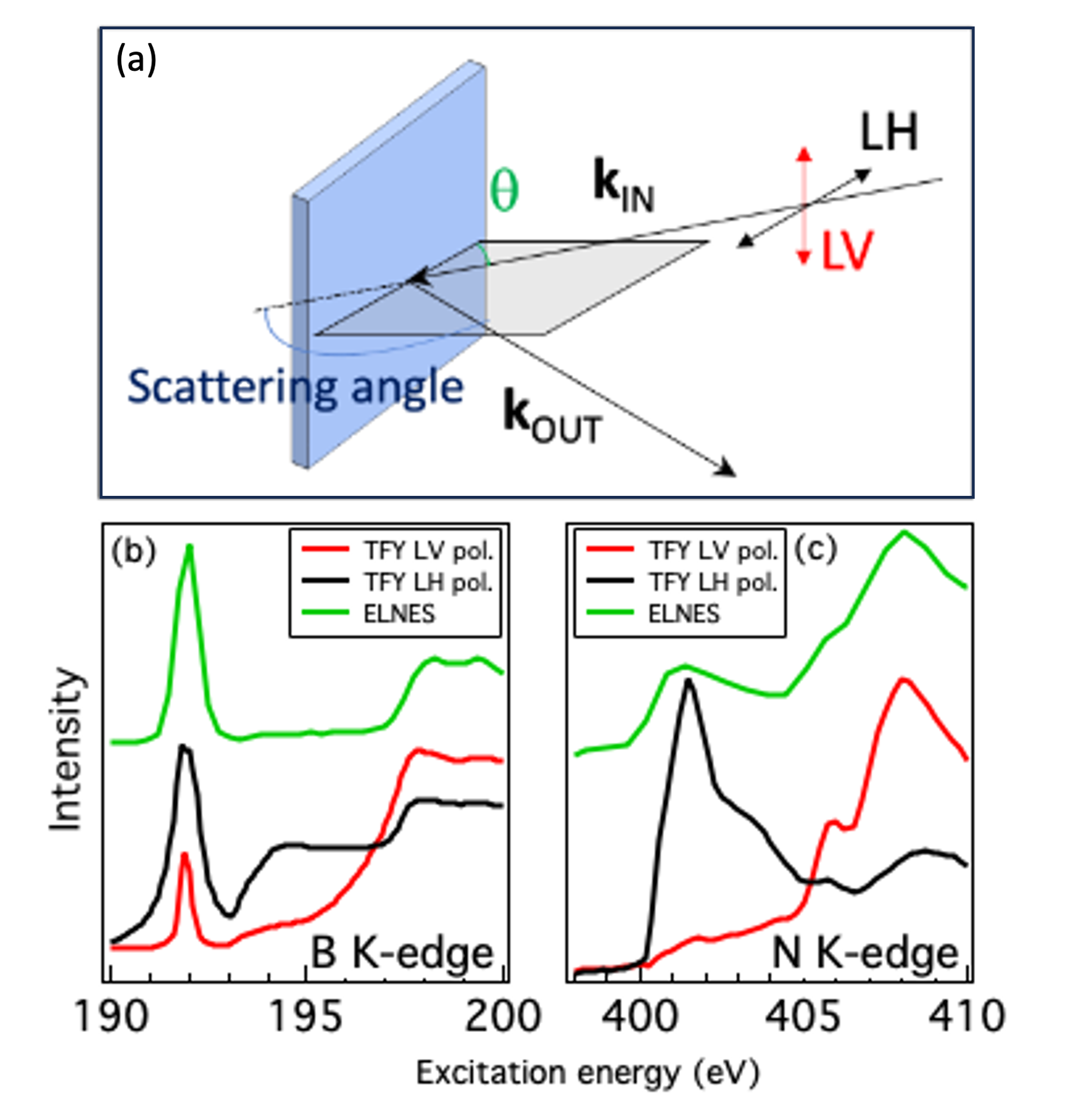}
			\caption{ \label{XAS} (a) Experimental geometry. (b) B K-edge XAS and ELNES and (c) N K-edge XAS and ELNES.  The XAS spectra are obtained from total fluorescence yield (TFY) measured in both LH and LV polarizations  
   at 30$^\circ$ and 20$^\circ$ grazing incident geometry in (a) and (b), respectively. ELNES spectra are vertically offset for clarity.  
		} 
	\end{figure}

Fig. \ref{XAS}(b) shows the ELNES and XAS spectra at the B K edge measured from total fluorescece yield (TFY) in both linear vertical (LV) and linear horizontal (LH) polarizations.
They are consistent with available XAS, XRS, and ELNES results  \cite{Brown1976,Davies1981,Chaiken1993,Jimenez1996,Franke1997,Carlisle1999,MacNaughton2005,Hua2012,Watanabe1996,Feng2008,Leapman1979,Arenal2007,Fossard2017}.
XAS spectra have a strong polarization dependence,  which is related to the crystal anisotropy of {\hBN}.
The LV polarization selects orbitals lying within the BN layers.
Since the incident angle is 30$^\circ$ grazing, i.e, 60$^\circ$ with respect to the $c$ axis,  
the LH polarization instead selects orbitals oriented both within the BN layers and perpendicular to it.
The sharp exciton peak\cite{Carlisle1999,Feng2008,Karsai2018} at 192 eV, which is more intense for LH polarization, indeed corresponds to a transition into  $\pi^*$ states that have distinct $p_z$ character (i.e., orthogonal to the BN layers), whereas spectral features above 198 eV are excitations into $\sigma^*$ states.
The fact that the $\pi^*$ peak at 192 eV is not completely suppressed in LV polarization is likely due to the fact that the surface of the sample is not completely flat and/or slightly tilted so that its normal is not exactly lying in the scattering plane. In any case, the absence of additional structures in the spectra and the strong dip  after the $\pi^*$ peak are an indication of the very good quality of the sample\cite{Caretti2011,Niibe2010,McDougall2017}.

The N K edge ELNES and XAS spectra, which also agree with results from literature \cite{Chaiken1993,Franke1997,MacNaughton2005,Miyata2002,Vinson2017,Watanabe1996,Arenal2007}, are displayed in Fig. \ref{XAS}(c).
As for the B K edge, also the XAS spectra remain strongly polarization dependent.
Their interpretation  is analogous to the  B K edge.
The first prominent peak at 401 eV, which is visible mainly for LH polarization, is due to an excitation into $\pi^*$ states.
Excitations into $\sigma^*$ states are responsible for the peaks at energies above 408 eV.

As expected (see the end of Sec. \ref{theory}), at both edges ELNES and XAS results are characterised by the same spectral
features. 
The remarkable agreement between the spectra obtained with the two techniques in Fig. \ref{XAS} additionally demonstrates that the energy calibration in the two measurements is fully consistent. The access to anisotropy effects is more difficult  with ELNES than with XAS. In fact, the field responsible for electronic excitations is parallel to the direction of momentum transfer and, therefore, at zero scattering angle, parallel to the direction of electron propagation. 
Therefore, ELNES spectra obtained for electron-beam incidence parallel or orthogonal to the {\hBN} planes could  be in principle compared with equivalent XAS measurements performed in vertical or horizontal polarisation, respectively. However, EELS is generally performed with focused electron probes and it integrates a finite-momentum range resulting in contributions for both in-plane and out-of-plane excitations.

\subsection{Resonant inelastic x-ray scattering spectra}
\label{sec:RIXS}

\begin{figure*}
	\includegraphics[width=1.4\columnwidth]{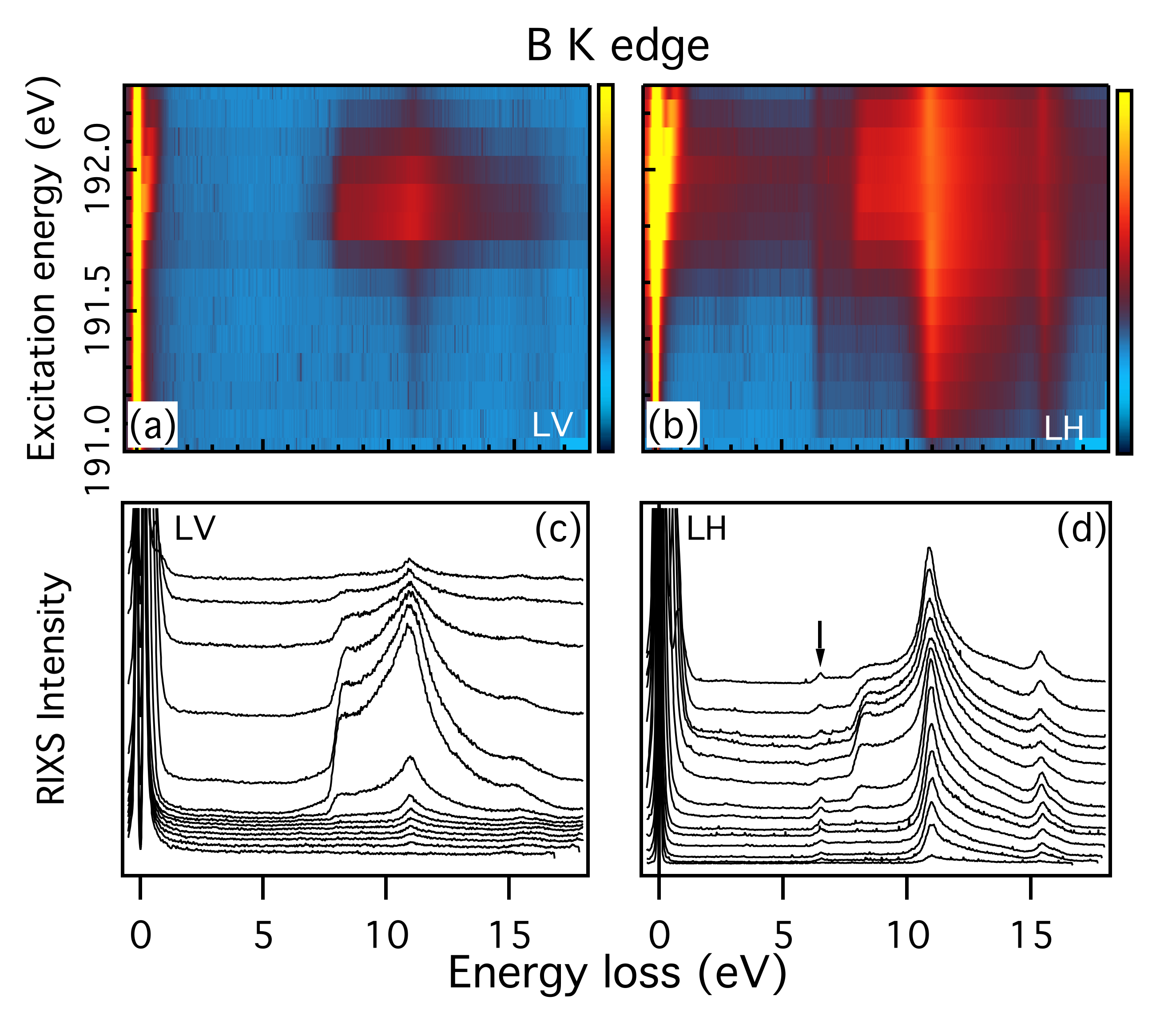}
\caption{ \label{RIXS_map_B}
			(a)-(b) B K-edge RIXS map in LV and LH polarizations respectively and (c) and (d) the corresponding stacks, for incident energies between 191 eV (bottom spectrum) and 192.3 eV (top spectrum) with steps of 0.1 eV. The arrow in panel (d) marks a Raman feature at 6.4 eV. }
\end{figure*}

In an ionic solid like {\hBN}, valence and conduction states are localised mainly on different atomic sites.
As a result, in XAS spectra, in which the conduction level is filled by the excited core electron, core excitons are generally more strongly bound at cation sites than at anion sites \cite{Pantelides1975}.
The RIXS process instead involves both  conduction states (in the photon absorption)
and valence states (in the photon emission). Therefore, the question that naturally arises is whether  it is more convenient to measure the RIXS spectra at the B or N K edge in order to characterise the main neutral electronic excitations in {\hBN}.

The top panels (a) and (b)  of Fig. \ref{RIXS_map_B} display the RIXS maps acquired at B K edge in both  LV  and LH polarizations. 
The maps are represented as a function of the excitation energy $\w_1$ (vertical axis) and energy loss $\w=\w_1-\w_2$ (horizontal axis).
The corresponding individual spectra are shown in Fig. \ref{RIXS_map_B} (c) and (d).
The chosen excitation energies $\w_1$, between 191 eV and 192.3 eV, scan
across the first $\pi^*$ exciton peak in XAS [see Fig. \ref{XAS}(a)].
We note that this resonance is very strong: the elastic line (at 0 eV energy loss) increases by a factor 200 between an incident photon energy of 191 eV and the $\pi^*$ excitation at 192 eV.

In RIXS the spectral features that are located at the same energy loss independently of the excitation energy are called Raman losses and  correspond to neutral excitations of the system.
In both polarizations, %at the B edge
the spectra show three main Raman features: 
a sharp step at 8 eV, a main peak at 11.5 eV and a second peak at 16 eV.
These high-energy excitations have been interpreted as different $\pi$-$\pi^*$ and $\sigma$-$\pi^*$ interband transitions \cite{Jia1996,Carlisle1999,MacNaughton2005,Fugallo2015}.

Most importantly, in LH polarization here we discover (see black arrow) an additional Raman excitation peaking at 6.4 eV energy loss %together with a weaker shoulder at 6.2 eV
that was missed in the previous RIXS studies on {\hBN} (probably, for the lower quality of the samples, crushed powders in Refs. \cite{Jia1996,Carlisle1999}, the lower resolution, 0.5 eV in Refs.\cite{Jia1996,Carlisle1999} and 0.3 eV in Ref. \cite{MacNaughton2005}, or the very long tail of the elastic peak in Ref.\cite{MacNaughton2005}).
In the following subsection we will analyse this low energy excitation in more detail.

\begin{figure*}
	\includegraphics[width=1.4\columnwidth]{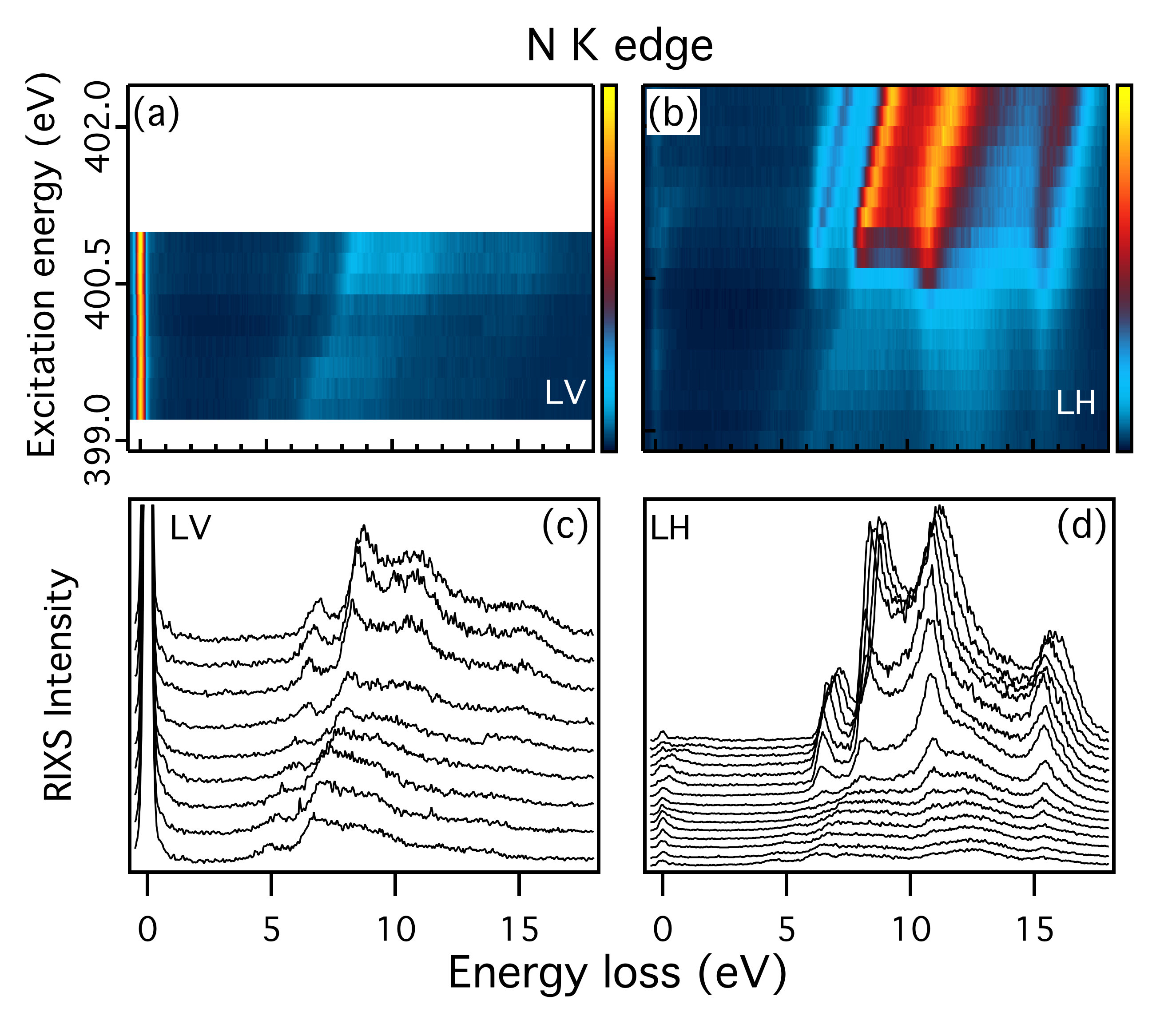}
\caption{ \label{RIXS_map_N}
			(a)-(b) N K-edge RIXS map in LV and LH polarizations respectively and (c) and (d) the corresponding stacks. The incident energies span between 399.3 eV (bottom spectrum) and 400.9 eV (top spectrum) for LV polarization and between  398.9 eV (bottom spectrum) and 402.3 eV (top spectrum) for LH polarization, with steps of 0.2 eV in both cases.}
\end{figure*}

Fig. \ref{RIXS_map_N} shows the RIXS maps [panels (a)-(b)] together with the individual spectra [panels (c)-(d)] measured across the N K edge in both  polarizations. 
In this case we observe a qualitatively different behaviour with respect to the B K edge RIXS spectra in Fig. \ref{RIXS_map_B}.
In the spectra at the N edge all the energies of the main peaks
have a linear dependence on the excitation energy. This implies that they are not Raman losses, but fluorescence features.
As in the normal N K edge emission, they can be interpreted in terms of the valence density of states, projected on the N atomic sites and bearing a $p$ angular component\cite{MacNaughton2005,Vinson2017}.
The dominant contribution of N $2p$ states is in the upper region of the valence band, in a range spanning 10 eV  (whereas N $2s$ are mainly located at the bottom of the valence band). 
Remarkably, the lowest-energy Raman loss at 6.4 eV found at the B edge in LH polarization is not clearly visible at the N edge, where the lowest-energy peak in both polarizations  does not show a pure Raman behavior.
The most plausible explanation\cite{Urquiza2024} for the absence of this Raman feature is that the lowest conduction states in {\hBN} have mainly B character and therefore cannot be reached exciting an electron at the N K edge.
Therefore, the important first  answer that we obtain from this survey of the RIXS results is that in the ionic {\hBN} compound the determination  of lowest neutral excitations is more favorable at the cation B K edge than at the anion N K edge. 

In the next subsection, we will relate the RIXS results at the B K edge with the spectra obtained by EELS and ellipsometry, and with {\it ab initio} BSE calculations.

\subsection{Connection with EELS and optical absorption}
\label{sec:rel_EELS_abs}

We may expect that RIXS, on one side, and  EELS (and NRIXS), on the other side, being very different spectroscopies, give rise to completely different spectra of neutral excitations in materials. 
Since the matrix elements [i.e., the squared moduli multiplying the delta functions in Eqs. \eqref{eq:RIXS}-\eqref{eq:NRIXS}] are  not the same, the two spectral shapes can be very different.
A notable example are $d$-$d$ transitions within the gap of transition metal compounds that are routinely measured by RIXS
at small $\bfq$, 
whereas they have negligible spectral weight in EELS at the same momentum transfer, since they are forbidden in the dipole limit  and become visible in the loss function only at large momentum transfers\cite{Larson2007,Hiraoka2009,Gloter2009,Ghiringhelli2005,Chiuzbaian2008,Huotari2008}.
However, a careful analysis of Eqs. \eqref{eq:RIXS}-\eqref{eq:sqw} suggests that RIXS and  EELS  can still measure the same electronic excitations: they share the same delta functions expressing energy conservation in Eqs. \eqref{eq:RIXS}-\eqref{eq:NRIXS}. 
Therefore, if RIXS and EELS spectra have peaks at the same energy, we can, in principle, use either technique to interpret the other\footnote{Extracting the dynamical structure factor directly from RIXS spectra instead requires additional hypotheses\cite{Brink2006,Ament2011}.}. 

A possible origin of further discrepancy in practice between measured  RIXS and EELS spectra is their respective momentum resolutions (see Secs. \ref{sec:exp_comp_details_RIXS}-\ref{sec:exp_comp_details_EELS}).
In EELS the momentum integration is much larger than in the soft X-ray RIXS measurements, even though the EELS cross section \eqref{eq:EELS} favors the small $\bfq\to0$  components.
Even taking into account these precautions, it remains sensible to directly compare EELS and RIXS spectra.

\begin{figure}[t]
	\includegraphics[width=0.49\textwidth]{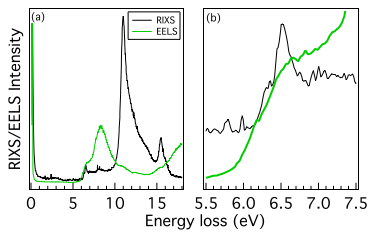}
			\caption{ \label{RIXSvsEELS} Comparison of  RIXS at B K edge in LH polarization (black lines) and  EELS (green lines) spectra. The right panel (b) shows a zoom of the same spectra on the energy range of their onset.
   }
	\end{figure}

Therefore, in order to establish  how we can benefit in practice from a fruitful combination of  EELS  and RIXS spectroscopies, we have also measured the EELS loss function of {\hBN} on the same type of samples used in RIXS experiments. 
The measured EELS spectrum is in agreement with previous {\hBN} results\cite{Fossard2017,Schuster2018} at small $q \lesssim 0.1$ \AA$^{-1}$. 
Fig. \ref{RIXSvsEELS} compares the B K edge RIXS spectrum  at 191.4 eV excitation energy in LH polarization 
with the EELS spectrum. The right panel (b) of Fig. \ref{RIXSvsEELS} shows a closeup of the same spectra in the energy range corresponding to the spectral onset. 
Also taking into account that the momentum transfer in RIXS  
is much smaller than the EELS angular integration, we find that both RIXS and EELS spectra in Fig. \ref{RIXSvsEELS} display the same features at 8 eV and 6.4 eV, with a less intense shoulder at 6.2 eV in RIXS  and at  6.17 eV in EELS.
 The main peak at 11.5 eV in the RIXS spectrum becomes in the EELS spectrum a high-energy shoulder of the 8 eV main peak.
From previous EELS and NRIXS investigations\cite{Mamy1981,Tarrio1989,Fossard2017,Galambosi2011,Fugallo2015}, we know that the 8 eV peak is the $\pi$ plasmon of {\hBN}. 
This very good matching between EELS and RIXS proves the capability of RIXS to investigate also collective charge excitations that are delocalised as the 
$\pi$ plasmon of {\hBN}. 
Instead, the origin of the common excitation peaking at 6.4 eV remains to be elucidated.

	\begin{figure}
	\includegraphics[angle=270,width=0.49\textwidth]{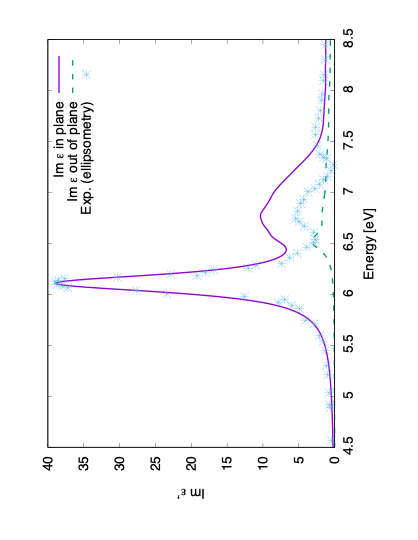}\\
	\includegraphics[angle=270,width=0.49\textwidth]{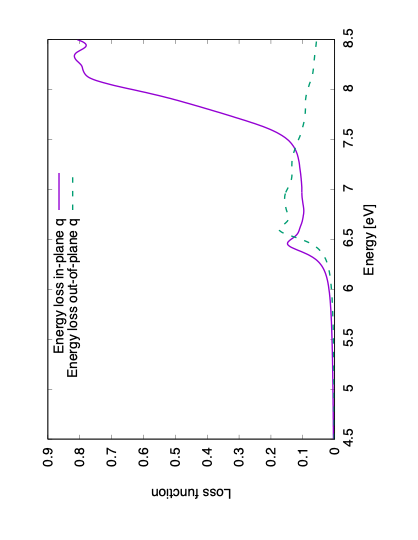}\\
	\includegraphics[angle=270,width=0.49\textwidth]{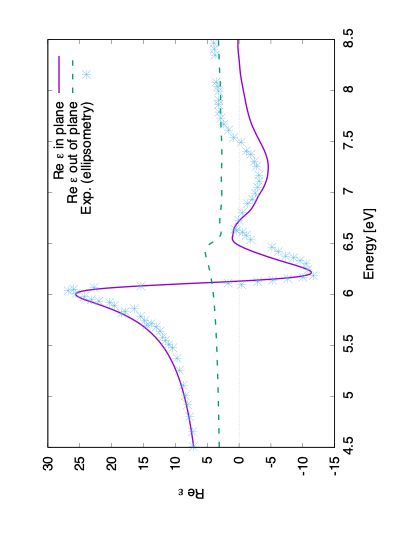}
			\caption{ \label{ellivsbse} BSE calculations and data from ellipsometry experiment from Ref. \onlinecite{Artus2021} for (upper panel) $\text{Im} \epsilon$, (middle panel) loss function $-\text{Im} \epsilon^{-1}$ and (bottom panel) $\text{Re} \epsilon$. The BSE calculations are done for $\bfq\to0$ in both in-plane (solid purple lines) and out-of-plane (dashed green lines) directions. Ellipsometry data (light blue crosses) are available only for in-plane polarization. }
	\end{figure}

To interpret RIXS results, RIXS spectra are often directly compared to optical absorption spectra (see e.g. \cite{Ishii2004,Kim2004,Grenier2005,Braicovich2007,Benckiser2013,Bachar2022,Vorwerk2020}). 
In the upper panel of Fig. \ref{ellivsbse}  we report the experimental absorption spectrum (light blue crosses), $\text{Im}\epsilon(\w,\bfq\to0)$, for in-plane polarization $\bfq\to0$,  from a recent accurate ellipsometry experiment\cite{Artus2021}.
The main exciton peak at the absorption onset is located at 6.12 eV, i.e., at a lower energy than the lowest energy features in the EELS and RIXS spectra in Fig. \ref{RIXSvsEELS}. This mismatch  evidences how the excitations measured by RIXS are actually {\it not} the same as those probed by optical absorption.

In the upper panel of Fig. \ref{ellivsbse}, the experimental absorption 
is also compared to the theoretical $\text{Im}\epsilon(\bfq\to0,\w)$ spectrum calculated from the solution of the BSE for both in-plane $\bfq\rightarrow0$ (solid line) and out-of-plane $\bfq\rightarrow0$ (dashed line). In this energy range, the out-of-plane spectrum has a much lower intensity, since $\pi$-$\pi^*$ transitions are dipole forbidden in the out-of-plane polarization.
As in previous studies\cite{Arnaud2006,Artus2021}, the agreement between theory and experiment for the optical absorption spectra is very good: the BSE is certainly reliable for the description of excitons in {\hBN}. 
Using the same formalism, we have also calculated  the loss function  $-\text{Im}\epsilon^{-1}(\bfq\to0,\w)$, see the middle panel of Fig. \ref{ellivsbse}.
Contrary to the absorption spectra in the upper panel of Fig. \ref{ellivsbse}, the loss functions in the two directions have similar intensities at their respective onsets.

The simulated spectra compare very well with the experimental EELS spectrum in Fig. \ref{RIXSvsEELS}, also considering the fact that while the calculation is done at $\bfq\rightarrow0$, the experimental spectrum is a result of an integration over a finite solid angle around $\bfq=0$. 
Most importantly, we find that, as in the EELS spectrum, in the calculated loss function for the in-plane $\bfq\to0$ the main plasmon peak at 8 eV is preceded by a smaller feature peaking at 6.45 eV.
This first peak in the loss function is located where also  $\text{Re}\epsilon=0$ (see the bottom panel of Fig. \ref{ellivsbse}), it is therefore by definition a longitudinal collective excitation\cite{Bechstedt2014}. 

However, contrary to the 8 eV peak, it is not a plasmon\cite{Koskelo2017,Wirtz2005}: its physical origin is  not the repulsive electron-hole exchange $\hat V$ in Eq. \eqref{eq:BSE} but the attractive electron-hole interaction $\hat W$. It is therefore a bound longitudinal exciton\cite{Agranovich1984}.
Indeed, it is located well inside the direct band gap of {\hBN} (6.95 eV in the GW calculation\cite{Artus2021}), while its energy is larger than the one of the transverse exciton measured by optical absorption (see the top panel of Fig. \ref{ellivsbse}). The energy difference between the two is called longitudinal-transverse splitting, and is due to the long-range component of the Coulomb interaction \cite{Agranovich1984,Galamic2005,Bechstedt2014}.
On the basis of the solution of the BSE we evaluate 
the longitudinal-transverse splitting to be 350 meV.
This value is smaller than the estimation of 420  meV in Ref. \onlinecite{Elias2021} that has been obtained from a fit of the experimental optical spectra.
The BSE longitudinal-transverse splitting in {\hBN} turns out to have a value similar to the analogous BSE calculations\cite{Galamic2005} in Ne and Ar (300 meV and 360 meV, respectively\footnote{We note that both BSE results in Ref. \onlinecite{Galamic2005} overestimated the corresponding experimental values.}). 

In summary, the direct comparison between RIXS and EELS spectra has shown that they have features in common, whereas 
excitations measured by RIXS and optical absorption are generally not the same.
Besides the $\pi$ plasmon at 8 eV,
our analysis has demonstrated that the lowest electronic excitation that both RIXS and EELS find in {\hBN} is a longitudinal exciton due to the same 
$\pi-\pi^*$ electron-hole pairs. 

It is worth noticing that these states are optically excited using in-plane polarization, whereas in RIXS, they are excited using LH polarization, which is orthogonal to the planes. This can be attributed to the fact that absorption at the K edge involving transitions to $p_z$ orbitals, can be achieved through out-of-plane polarization, as discussed in Sec. \ref{sec:XAS}.
In the next subsection, we investigate in detail the dispersion of this exciton as a function of the wavevector $\bfq$.

\subsection{Exciton dispersion}
\label{sec:exc_disp}

To measure the exciton dispersion, we have carried out RIXS measurements as a function of the incident angle, using LH polarization with an excitation energy $\w_1=190.4$ eV.
This corresponds to an experimental configuration where the exciton is clearly visible. 
The in-plane momentum transfer $q_\parallel$ changes between $-0.01$ and $-0.07$  \AA$^{-1}$ along the $\Gamma$M  direction of the Brillouin zone and between $-0.04$ and $0.07$ \AA$^{-1}$  along $\Gamma$K (see Tab. \ref{tabq}).
The out-of-plane component  $q_\perp$ varies concomitantly between 0.11 and 0.13 \AA$^{-1}$, but its effect on the exciton dispersion can be considered negligible due to the layered nature of \hBN \cite{Galambosi2011,Fugallo2015,Fossard2017}.

\begin{figure}[ht]
		\includegraphics[width=1\columnwidth]{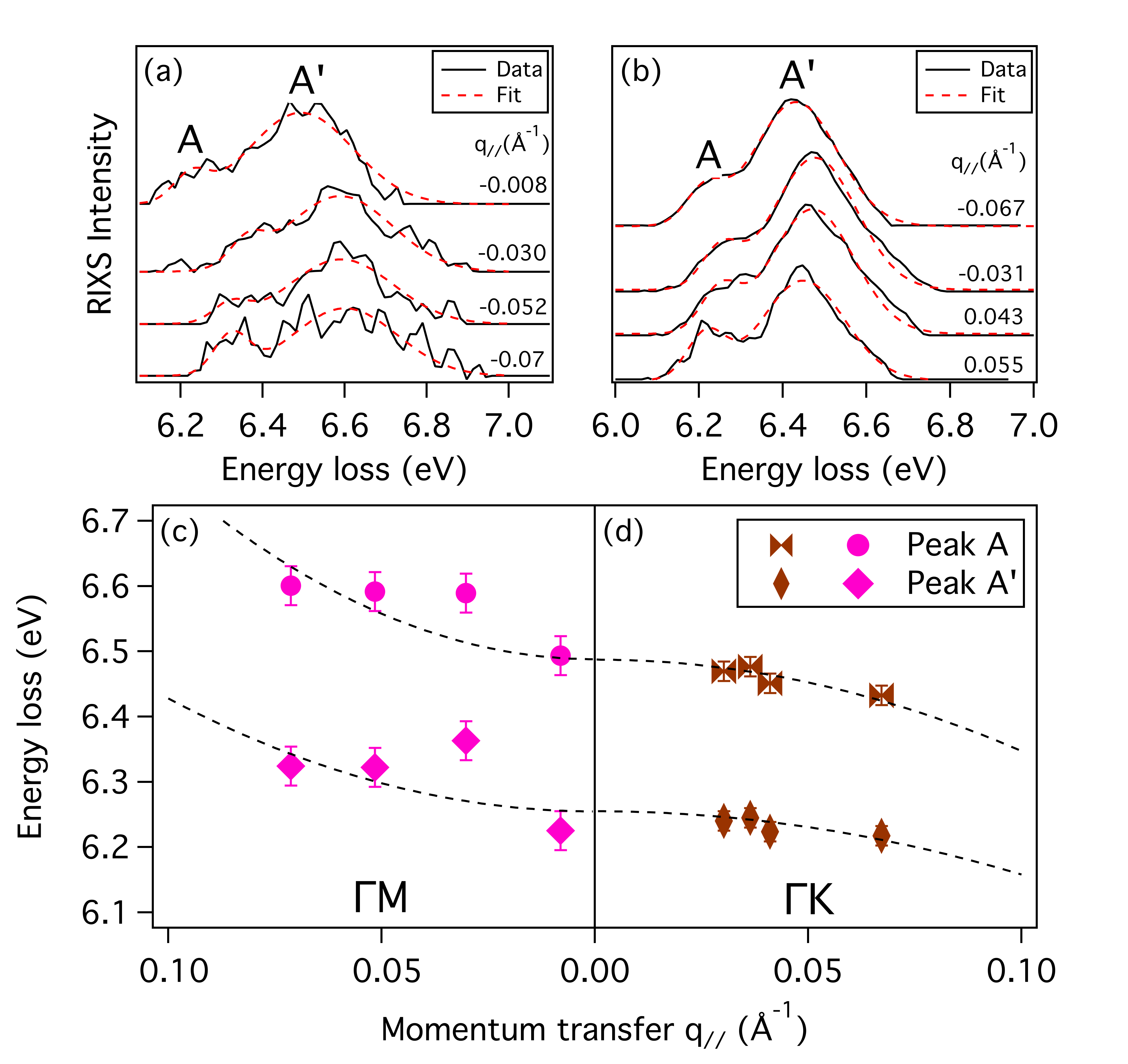}
		\caption{ \label{ExcitonDispersion} (a)-(b) RIXS spectra acquired in LH polarization as a function of the incident angle along $\Gamma$M and $\Gamma$K respectively. Values of in-plane momentum transfer $q_\parallel$ are indicated aside each spectrum (c) Exciton dispersion along (left panel) $\Gamma$M and (rigth panel) $\Gamma$K: energies of the maxima of the excitonic peaks A and A' as a function of $|q_\parallel|$. %The inset displays the experimental geometry. %\matteo{-0.05}
		}
	\end{figure}

The features corresponding to the two lowest-energy excitations at 6.2 and 6.4 eV, marked A and A', have been extracted from the rest of the RIXS signal by subtracting a linear background under the two peaks. They give rise to the spectra shown in Fig. \ref{ExcitonDispersion}(a) 
and Fig. \ref{ExcitonDispersion}(b), for momentum transfers $q_\parallel$ along $\Gamma$M and $\Gamma$K, respectively.
The spectra have been fitted by means of two gaussian functions. All the fit parameters were left free for the $\Gamma$K direction. The FWHM was found to be sharper for the A feature with respect to the A' one, being respectively about 0.1 eV and 0.2 eV. Due to the poorer statistics of the data along the $\Gamma$M direction, here the fit was done by keeping  the FWHM of A and A' fixed to 0.1 eV and 0.3 eV respectively.

\begin{table}[ht!]
\caption{In-plane $q_\parallel$ and out-of-plane $q_\perp$ components of momentum transfers in the RIXS experiment (see Fig. \ref{ExcitonDispersion}).\\
\\
\label{tabq}} 
\begin{ruledtabular}
\begin{tabular}{c c c}
   Direction & $q_\parallel$ (\AA$^{-1}$)  & $q_\perp$ (\AA$^{-1}$) \\
  \hline
   $\Gamma$M & -0.07 & 0.11 \\
             & -0.05 & 0.12 \\
             & -0.03 & 0.13 \\
             & -0.01 & 0.13 \\
  \hline
   $\Gamma$K & -0.04 & 0.12 \\
             & -0.03 & 0.13 \\
             & 0.04  & 0.12 \\
             & 0.07  & 0.11 \\
\end{tabular}
\end{ruledtabular}
\end{table}
	
	Fig.  \ref{ExcitonDispersion}(c) summarizes the result of the fits, displaying the measured excitation energies as a function of the absolute value of the momentum transfer $|q_\parallel|$.
	The two features A and A' follow parallel parabolic-like dispersions along both directions.
    However, their dispersion is positive along $\Gamma$M, while it is negative along $\Gamma$K: the largest energy is at $\bfq=\Gamma$ and decreases by increasing $|q_\parallel|$. 
    This negative dispersion is due to the fact that {\hBN} has an indirect band gap\cite{Arnaud2006} along $\Gamma$K. As a consequence, increasing the momentum transfer $|q_\parallel|$, the smallest energy separation between the bottom conduction and top valence band decreases. 
    The present results are complementary to the EELS findings in Ref. \onlinecite{Schuster2018}, where the negative exciton dispersion was measured in the loss function for larger momentum transfers $q>0.1$ {\AA} along $\Gamma$K.  
    Here, soft X-ray RIXS demonstrates to be a powerful way to identify the nature of the band gap in semiconductors and insulators.

\section{Discussion}

\subsection{Analogy to orbital excitations in transition metal compounds}
\label{sec:analogy}

Transition metal compounds, where the electrons are tightly localised around the atoms, are often considered as paradigmatic strongly correlated materials\cite{Mott1990,Fujimori1992,Imada1998}.
In these materials, transition metal ions have  partially filled $d$ shells:
neutral electronic excitations can therefore take the form of orbital $d$-$d$ excitations localised at the transition metal ions, in which one (or more) electron is promoted from an occupied $d$ orbital to another, empty, $d$ orbital.
The details of these local $d$-$d$ transitions are  well described by cluster models  with a single ion in a crystal field \cite{vanVeenendaal2008,DeGroot2008,Haverkort2012}.
Instead, orbital excitations that have an appreciable dispersion as a consequence of the superexchange interaction, which makes them mobile, are called orbitons\cite{Kugel1982,vandenBrink1998,Khaliullin2005} (or orbital waves). By definition, these dispersive excitations cannot be captured by finite clusters: their description requires the introduction of specific spin-orbital models\cite{Kugel1982,vandenBrink1998,Khaliullin2005}.
Like magnons in a magnetic system, orbitons are the collective excitations of the orbital degrees of freedom. %From a solid-state physics point of view, instead, 
Alternatively, orbital excitations, whose peaks in the spectra are located well within the photoemission band gaps, can be considered as strongly bound excitons\cite{Lee2010}. In this sense, they are similar to localised Frenkel excitons in wide-gap insulators or molecular solids\cite{Knox1963,Bassani1975}. 
Their dispersion can be generally understood in terms of  electron and hole hoppings (i.e., the band dispersion) and/or dipole-dipole electron-hole exchange interactions\cite{Cudazzo2015}.

This general connection allows us to suggest a parallelism also with the exciton in {\hBN}, a material that definitely has different physical properties than transition metal compounds and, obviously, is not considered strongly correlated (notably, {\hBN}  is not magnetic and superexchange interactions are not at play).
Indeed, analogously to a $d$-$d$ excitation at a transition-metal ion, in {\hBN} the exciton corresponds to a molecular $\pi$-$\pi^*$ orbital excitation, situated on a BN molecule, and is tightly bound (i.e., its excitation energy is well below the band gap). 
Moreover, it is a collective longitudinal excitation (corresponding to a zero of $\text{Re}\epsilon$) with a non-negligible dispersion, suggesting that it could  alternatively be termed with the evocative name  of ``$\pi$-orbiton''.
Along this line, one could also note that, in analogy to cuprates\cite{Schlappa2012}, {\hBN} showcases an orbital-charge separation\footnote{Contrary to cuprates, in {\hBN} the spin degree of freedom is, of course, inactive.}, evidenced by the separation between the $\pi$-$\pi^*$ exciton at the spectral onset and the higher-energy $\pi$ plasmon, both associated with the same $\pi$ electrons.

\section{Conclusion}
\label{sec:conclusion}

In summary, we have characterised the lowest energy exciton in {\hBN} by means of RIXS, identifying spectral features that were missed in previous measurements.
We have determined the exciton dispersion along the -K$\Gamma$K and $\Gamma$M directions for $q<$ 0.1 \AA$^{-1}$, complementing available results for the loss function\cite{Schuster2018} for $q>$ 0.1 \AA$^{-1}$. 
We have contrasted  the positive exciton dispersion along $\Gamma$M to the negative dispersion along $\Gamma$K, which is a consequence of the indirect band gap in {\hBN}. 
This also demonstrates that soft x-ray RIXS is a powerful tool to measure the dispersion at small $q$ of excitons in low-dimensional materials.
Finally, we have directly compared RIXS with EELS, showing that they probe the same longitudinal exciton, at variance with absorption spectra where is the transverse exciton that is observed. %This analysis has also allowed us to estimate the exciton longitudinal-transverse splitting in {\hBN}, equal to ??? 
The combination of different spectroscopy techniques, such as RIXS and EELS, and {\it ab initio} calculations is an insightful and rewarding strategy to characterise electronic excitations in materials.

\appendix
\section{Low-energy excitations at the N K edge}

  \begin{figure}[ht]
	\includegraphics[width=\columnwidth]{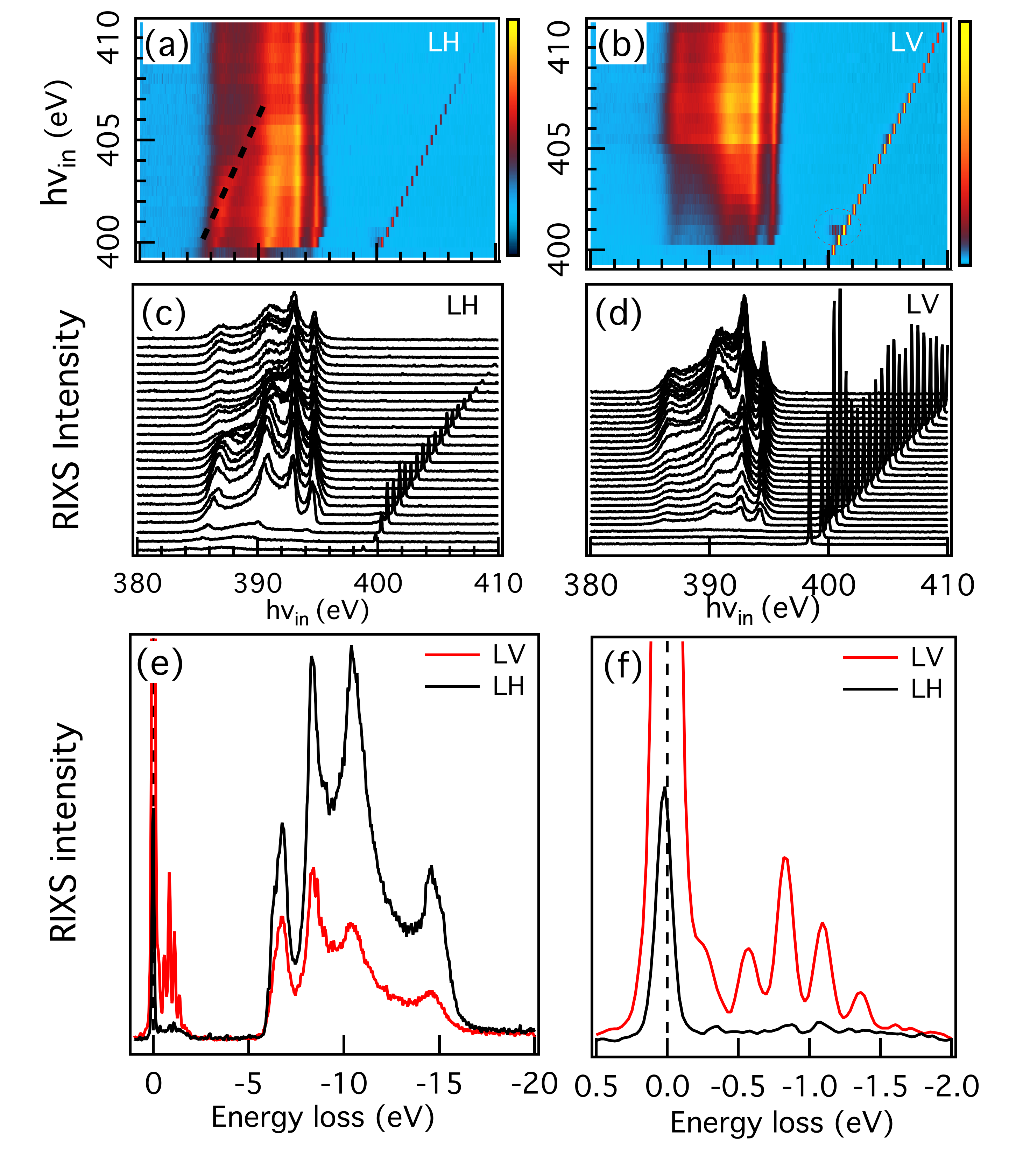}
	\caption{ \label{RIXS_NK_LowELoss}(a)-(b) Low energy N K-edge RIXS maps, as a function of incident photon energy (vertical axis) and emitted photon energy (horizontal axis),  in LH and LV polarizations, respectively, and (c)-(d) the corresponding stacks of individual spectra. (e)-(f) RIXS spectra as a function of energy loss, acquired in LH and LV polarization using an excitation energy of 401 eV.}
   \end{figure}
%0.4\textwidth
	\begin{figure}[ht]
	\includegraphics[width=0.43\textwidth]{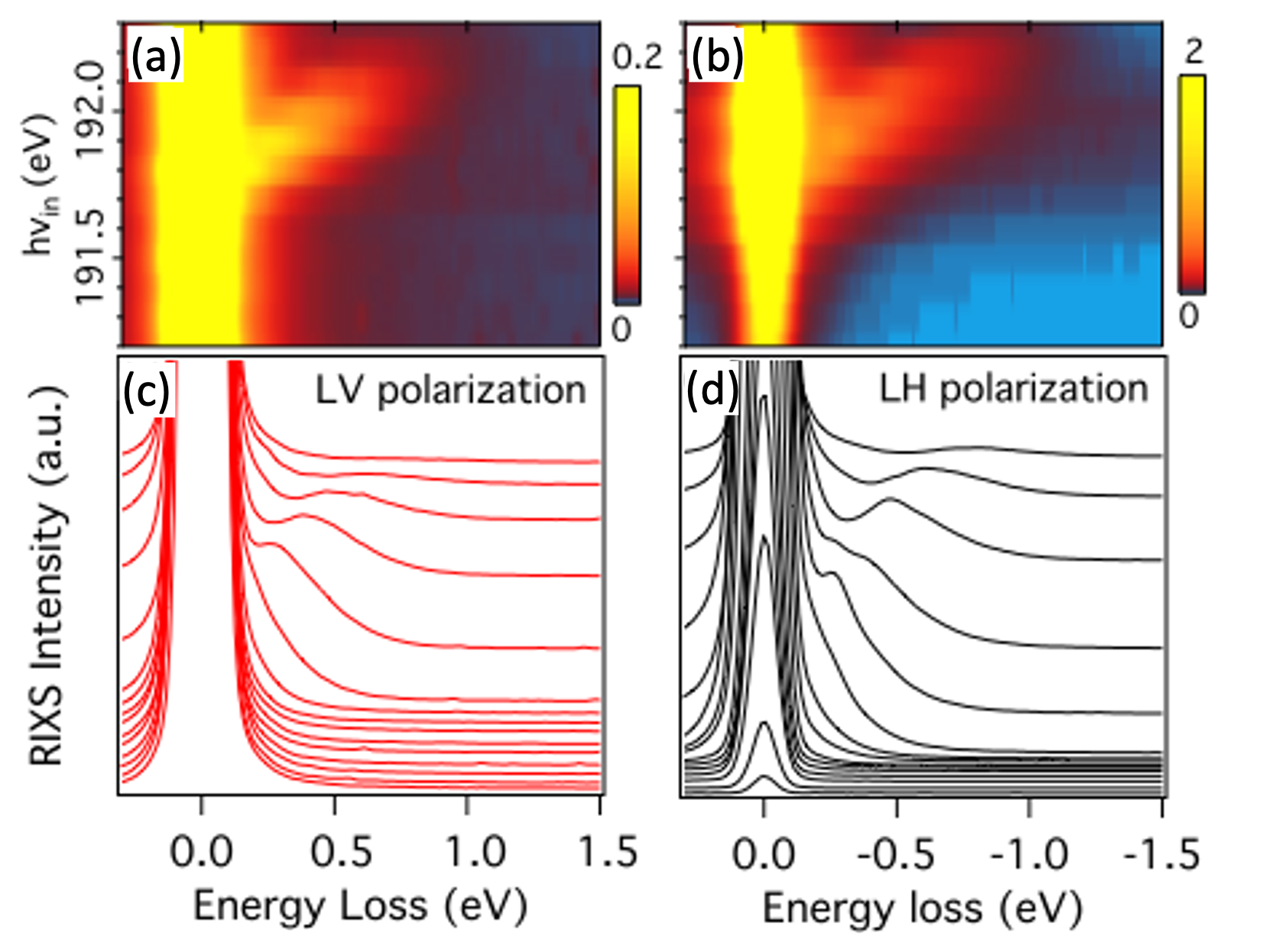}
	
			\caption{ \label{RIXS_BK_LowELoss} (a)-(b) Low-energy B K-edge RIXS maps, as a function of incident photon energy (vertical axis) and energy loss (horizontal axis), in LV and LH polarizations, respectively, and (c)-(d) the corresponding stacks of individual spectra. }
	\end{figure}
 
 In the RIXS spectra at the N K edge, in the low-energy region close to the elastic line,  clear vibrational modes of 270 meV energy are also visible for LV polarization, but not for LH polarization, see Fig. \ref{RIXS_NK_LowELoss}.
 These excitations disappear upon cleaving, so we attribute them to molecular N$_2$  or other impurities possibly adsorbed at the surface.
 However, some low energy features remain visible also in the cleaved samples at the B K edge, as shown by Fig. \ref{RIXS_BK_LowELoss}. In this case, they behave as a fluorescence. Moreover, they strongly resonate at the main XAS peak at 192 eV, they have a maximum at the constant emitted energy 191.6 eV, and they disappear in less than 1 eV. The most plausible origin of these low-energy excitations are defects or impurities. 
 Very recently, Ref. \cite{Pelliciari2024}   has similarly identified peaks in N K edge RIXS spectra of defective {\hBN} at multiples of 285 meV energy, correlated with
single-photon emitters.\\
\\\\

\begin{acknowledgements}
This work was supported by a public grant overseen by the French National Research Agency (ANR) as part of the ``Investissements d’Avenir'' program (Labex NanoSaclay, reference: ANR-10-LABX-0035), by the Magnus Ehrnrooth Foundation, by the Labex Palm (Grant No. ANR-10-LABX-0039-PALM) and by the BONASPES project (ANR-19-CE30-0007). The research leading to these results has received funding also from the People Programme (Marie Curie Actions) of the European Union’s Seventh Framework Programme (FP7/2007-2013) under REA grant agreement n. PCOFUND-GA-2013-609102, through the PRESTIGE programme coordinated by Campus France. Computational time was granted by GENCI (Project No. 544). K.W. and T.T. acknowledge support from the JSPS KAKENHI (Grant Numbers 21H05233 and 23H02052) , the CREST (JPMJCR24A5), JST and World Premier International Research Center Initiative (WPI), MEXT, Japan. Finally, this work is part of the IMPRESS project that has received funding from the HORIZON EUROPE framework program for research and innovation under grant agreement n. 101094299.
\end{acknowledgements}

\bibliography{Bibliography.bib}

\end{document}